# A Review on Cybersecurity in Smart Local Energy Systems: Requirements, Challenges, and Standards

Siyuan Dong*[a], Jun Cao[b], Zhong Fan[a]

[a]*School of Computer and Mathematics, Keele University, Keele, Newcastle ST5 5BG*

[b]*School of Geography, Geology and the Environment, Keele University, Keele, Newcastle ST5 5BG*

**Corresponding author: s.dong@keele.ac.uk (Siyuan Dong)*

## Abstract

Smart local energy system (SLES) is considered as a promising pathway facilitating a more effective and localised operation, benefited from the complex information and communication technology (ICT) infrastructures and Internet of things (IoT) technologies. As a part of the critical infrastructure, it is important to not only put effective detection and management to tackle potential cybersecurity issues, but also require considerable numbers of standards to ensure the security of the internet of things system to minimise the risks. This study aims to review the existing standards, investigate how the compatibility with SLES development, and identify the area to focus on in the future. Although existing standards and protocols are highly fragmented, our findings suggest that many of them can meet the requirements of the applications and infrastructures of SLES. Additionally, many standards have been introduced to protect information security and personal privacy due to their increasing importance. The research also suggests that the industry needs to produce more affordable and cyber-secured devices and services. For the government and regulators, relevant guidelines on the minimum function and security requirements for applications should be provided. Additionally, compliance testing and certifications should be in place and carried out by an independent third party to ensure the components of SLES ecosystem with a satisfied security level by design.

(Word Count: 6230)

**List of Abbreviations**

| | | | |
|---|---|---|---|
| *SLES* | Smart Local Energy System | *PLC* | Programmable Logic Controllers |
| *DER* | Distributed Energy Resource | *MTU* | Master Terminal Unit |
| *ICT* | Information and Communication Technology | *HMI* | Human-Machine Interfaces |
| *IoT* | Internet of Things | *AMI* | Advanced Metering Infrastructure |
| *SCADA* | Supervisory Control and Data Acquisition | *HAN* | Home Area Network |
| *RTU* | Remote Terminal Units | *WAN* | Wide Area Network |

# 1. Introduction

In the past few years, distributed energy resources (DERs) have been widely adopted due to the collective goal in tackling climate change. They emerged as a promising solution to reduce the carbon emissions and help the transmission of existing energy industry towards a cleaner and more decentralised manner. The proportion of the renewables has significantly increased in the total energy production mix in the UK in the recent years [1], such as offshore wind. Similar to most countries, the UK's energy system is based upon the "supplier hub model" [2], which means the electricity is generated from energy suppliers, transmitted via transmission and distribution networks, and finally consumed by end-users. The newly built large-scale DER sites are usually far away from the consumer end. Together with increasing energy demand, they pose new challenges to the existing energy system, which may require additional expensive generation assets and network reinforcement and expansion [3]. However, there is an alternative approach, solving the problem at the near-consumer side.

The demand for more active energy management and more affordable energy supply have contributed to the rapid growth in DER deployments at the near-user end [4]. The network operators must seek for new solutions to the problems and challenges brought by the increasing penetration of DERs. Meanwhile, they also need to make the best use of the enormous flexibility that potentially benefits the whole system. For the better integration, some potential solutions have been proposed to embrace the idea of localising energy supply and also provide additional flexibility and resilience, such as virtual power plants [5], local energy market [6], and aggregators [7].

Amongst all potential solutions, SLES is considered as a promising pathway facilitating a more effective and localised operation. Previous study [8] identified the benefits of SLES, including effective provision of energy service, enabling flexibility within and across energy vectors, improved resilience and ability to cope with failure, etc. SLES substantially benefits from its complex information and communication technology (ICT) infrastructures that can provide enhanced observability and distributed control on DERs. The smart elements, consisted of various IoT technologies, have enhanced the interoperation of the grid system by providing multi-directional information flow with adequate data from users, substations, transmission and generation sides [9]. These smart devices contribute to the provision of a real-time balance, monitoring, and control at high granularity and accuracy.

However, there is an outstanding concern regarding the security, because of the competing interests of different parties or stakeholders, high level of interdependence, and social complexity [10]. A study suggested that existing standards and guidelines have not provided any clear definition of roles that different parties play, and a common understanding of key security requirements is yet to be shared [11]. Additionally, the inherent vulnerabilities may potentially expose the system to potential attacks [12], because the controlling and monitoring is undertaken based on internet-protocols and off-the-shelf solutions. Similar to smart grid, the nature of SLES can be considered as a part of critical infrastructures, which will likely draw unwanted attention and become the target of cyber-attack. It therefore is vital to undertake thorough examination on the components and identify existing vulnerabilities to ensure the main security objectives are met. In order to protect the IoT from the potential external cyberattacks, it requires not only effective threat detection and management, but also a considerable number of well-designed standards are necessary to ensure the security of the IoT system to minimise the risks. It is therefore worth investigating and reviewing the currently existing standards and identify the area to focus on in the future.

The rest of the paper is structured as follows: in Section 2, the background of SLES is described, including its key components, features and benefits, and potential challenges and risks; Section 3 introduces general cybersecurity objectives and requirements for SLES and general energy system, and also reviews the existing standards related to cybersecurity; Section 4 discusses and explains the main findings from the review and proposes several suggestions for SLES planning and development; and conclusions and future works are summarised in Section 5.

## 2. Background of SLES

In general, there are many key components comprising an energy system, including production, conversion, transmission, distribution and consumption [13]. This structure also works for a small energy system at a local level, such as a community or a building, and SLES. SLES can transmit electric and information flow during the operation. The electric flow starts from energy producers and finish at end users in the traditional system. However, in SLES, different shareholders and components communicate through a bi-directional communication flow, with the assistance of substantial number of sensors, actuators other smart objects. Although there has not been any clear definitions or explicit frameworks for SLES, some key elements and functions have been identified in the previous study [8].

### 2.1. Key Components of SLES

In order to deliver features described previously, a localised and highly automated system infrastructure is required. Many components are connected to the SLES for operating, monitoring and controlling electricity flow and measurements. The SLES requires involvement of different stakeholders comprising of various domains, such as service provider domain, communication network, grid domain, advanced metering infrastructure and customer domains, which are similar to the smart grid structure defined in [14]. The SLES is a desired solution to deliver more interactive operation and localised energy supply and control, which will be increasingly challenging for the existing system setup. The existing cybersecurity techniques and standards may no longer meet the requirements of the SLES. Therefore, a SLES has different objectives and features to provide reliable communication architecture and power supply. Here are the key components of a SLES:

**Grid Domain** is a critical part of the SLES, managing the bulk energy generation and distribution. In the grid domain, a Supervisory Control and Data Acquisition (SCADA) system plays an important role, which is a type of industrial control system that can monitor and control assets over large geographical areas with the help of control equipment. The decentralised automation management and remote control are helpful to ensure the reliability of power supply and lower the maintenance costs of the network. There are four main parts in a typical SCADA system [15], a) data interface appliance like remote terminal units (RTUs) and programmable logic controllers (PLCs); b) communication network; c) central master terminal unit (MTU); and d) human-machine interfaces (HMIs). RTU is extremely important in SCADA system connecting many sensors and metres, which are responsible for collecting information and proceed with commands sent from central MTU. The MTU communicates with RTUs through a secured communication network to perform control, alarming and other operations.

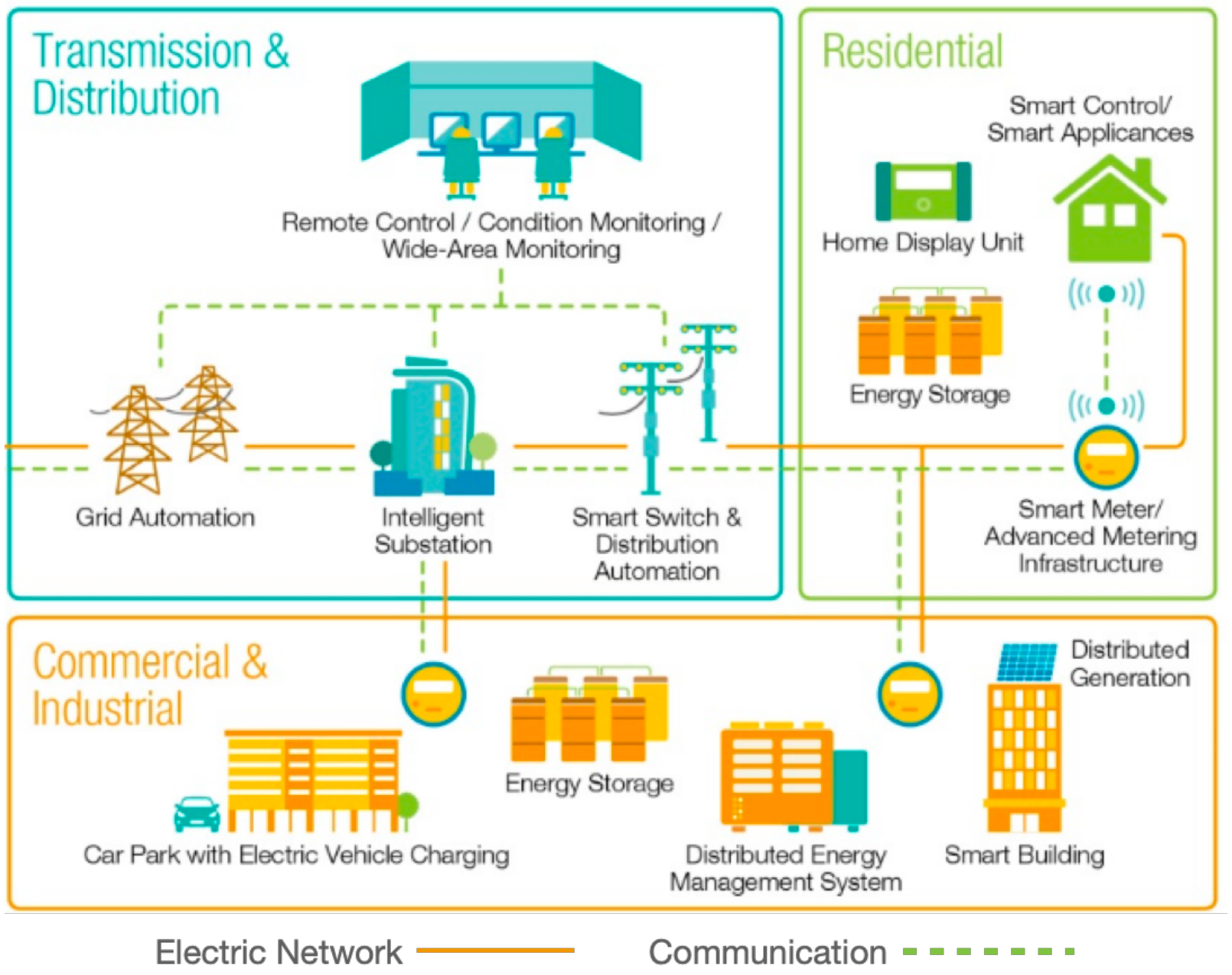

Figure 1 An potential layout of SLES adapted from [16]

**Advanced Metering Infrastructure (AMI)** plays a vital role in the SLES, acting as the connection between control centre and meters. It consists of home-area (HAN) and wide-area communication network (WAN), smart meter and data concentrators, and metre data management system [17]. The bi-directional communication between the central system and smart meters becomes easier due to the increasing penetration of IoT based technologies. The increasing installation of smart meter technologies enables the distribution networks to capture precise consumers' electricity usage along with other information. In this way, smart meters can collect and transmit data back to utility operators for better understanding of the energy consumption pattens. Smart meters can action upon request or in response to some events to the utility. In addition, smart meters are also found to be helpful for consumers to understand and improve energy consumption [18].

**IoT-based Communication Network** is considered one of the fundamental elements in the SLES. It enables the interaction between service provide domain and customer domain. The development of SLES is substantially contributed by IoT technologies that enable data flowing through the networks. Within the communication network, the utilisation of standard communication protocol enables each device or object to be individually addressed, and the communication between them can be near-real time. Therefore, different devices can be sensed and controlled remotely via a scalable communication network, enabling further integration of physical grid devices and computer-based control system. As the result, the IoT-based network can make system control and operation more efficient and accurate [19]. On the other hand, the increasing adoption of IoT devices also poses a challenge on the system with regards to the physical and cyber security of the infrastructure [20], which requires more complicated regulatory and technical measures.

**Customer Domain** includes smart appliances, premises networks and distributed energy resources. DERs are becoming increasingly popular in recent years, as they can efficiently provide end users more localised and cost-effective energy supply with technologies such as PV, wind turbine in tandem with energy storage. The consumers can manage the operation of DERs and smart appliances through the premise's networks, either HAN or WAN. However, the introduction of DERs results in bi-directional electricity flow, which creates some issues for the distribution network operators. The problem can be solved in an active approach by modifying the SCADA system to reconfigure distribution network based on changes in power flow. It requires a substantial amount of data from installed sensors that monitor system conditions such as faults and status of switches, significantly facilitated by the IoT-based network. Such an integration with the external power grid can enhance the active energy management from the consumer end and add greater reliability and resilience to the system.

**Service Provider Domain** includes markets, operators and service providers. It uses the communication network to coordinates the functions of energy generators, distribution and transmission network operators and consumers. The market creates a platform for all actors to participate and maintain the energy balance between the supply and demand. The operators ensure the delivery of service through the energy network provided by service providers. In the SLES, there may be more diversified actors in the service provider domain, as the results of localised governance and regulatory. Local authorities can take more responsibilities to govern the operation and the consumers may more actively participate in local energy market, where the local energy network operators will play a more transformative roles as a local system operator.

## 2.2. Key Features and Benefits of SLES

Similar to other local systems with DERs, a SLES aims to achieve local balancing, which means that it maximises the utilisation of energy produced locally to reduce the consumers' energy import from the external power grid [21]. In addition to electricity consumption, it is also able to integrate other services, such as heating and transport, to enable the local balancing across different sectors. The smart technologies are the fundamental attributes to realising more locality and flexibility in a SLES system. Smart elements are able to contribute to more accurate measurements and more digitalised and interactive management, and ultimately more autonomous operation. It enables the system to manage bi-directional information and electricity flows automatically, and therefore to provide greater capacity and resilience. For consumers, the smart elements can also provide consumers richer data and help them with better decision-making [22]. Therefore, these intelligent elements can make the best use of the information to optimise the system control and hence maximise the benefits for all stakeholders.

The other important feature of SLES is the localness. A SLES enables a more localised system management, operation and government based on their needs and system setup. Different from the traditional system, the ownership of a SLES can be more flexible and diverse, which may encourage more active participation and engagement of the local authorities, network operators and consumers [8]. Therefore, the SLES is very helpful to deliver more affordable energy and a fairer energy system. Additionally, local decision-making process will also make the service providers put more focus on consumers and quality of service, which can provide local customers an easy access to the system and address the desire to tackle the climate change locally. In this way, the locality of the SLES can not only help us exploit the value of system better, but also provide a location-specific solution to the energy transition.

*Table 1 Comparison between traditional grid and SLES* [8,20]

| Parameters | Existing Grid | SLES |
| --- | --- | --- |
| **Generation** | Centralised | Decentralised |
| **Communication** | None/one way | Two-way, real-time |
| **Customer Interaction** | Limited | Extensive |
| **Metering** | Electro-mechanical meters | Digital meters |
| **Operation** | Manual equipment checks and maintenance | Remote monitoring, predictive, time-based maintenance |
| **Maintenance** | Network operated by centralised network operators | Local authorities actively manage the SLES operation. |
| **Power Flow Control** | Limited | Comprehensive and automated |
| **Reliability** | Prone to failures and cascading outages | Automated, prevents outages before they start |
| **Restoration Following Disturbance** | Manual | Self-healing |
| **System Topology** | Radial. Power flows one-way | Network. Power flows multiple paths |
| **Distributed Generation** | Limited grid accessibility | Full and efficient grid accessibility |

## 2.3. Challenges and Potential Risks

As mentioned previously, a lack of understanding in SLES's framework and operation, hinder the development of SLES. However, it has some similarities to smart cities. For example, both are part of the key public infrastructure and both heavily reliant upon the participation of private companies and consumers. In traditional systems, the utility companies usually have the ownership of the entire infrastructures or utilise a managed service. Therefore, the cybersecurity may be prioritised during system acquisition so that the correct security measures can be in place. The emerging technologies, especially built upon IoT, are usually designed for the easy adoption so that the consumers can operate the devices through HAN or WAN. For this reason, the number of consumer-owned smart devices can easily exceed those owned and operated by the utilities. However, most consumers may not have the technical expertise or incentives to prioritise or maintain the infrastructure security. The divided administration makes the utility or system operators hard to monitor and manage devices, leading to a disparity in security protection. Therefore, administrative boundaries must be broken by interconnecting different networks to ensure the utility companies can operate smart devices and DERs together with consumers in a collaborative manner.

Most IoT-based devices adopted in the SLES are manufactured by third parties or private companies. Lu et al. [23] acknowledged that the secure operation of the power system is based on a stable ICT supply chain, and any disruption on its components can lead to unwanted and serious disruptive impacts on the whole system. Another study [24] also suggests that many security concerns and incidents can trace back to the inadequate management and risk of manufacturers and suppliers. In most of the real-life deployments, third parties or private companies are given access to key infrastructure assets and critical information without thorough reviews. Although such arrangement may help the developer deliver service faster and integrate with existing systems more easily, this may lead to catastrophic impacts if without proper management [25].

In addition, the utilisation of HAN and WAN provide an easy and flexible access to the system management. It enables users and utility companies to obtain more accurate consumer demand and status of DER production in higher resolution and participate in more complex system operation, such as demand side response. However, exposing the system to the external WAN may also increase the attack surfaces, leading to private data breach, device compromise, and even instability of the whole system [26]. The substantial growth in smart appliances and DERs in the SLES will essentially increase the cyber-physical interdependencies. The operation therefore will no longer merely depend upon the secure physical status of the infrastructure, bringing the importance of cybersecurity to an unprecedented level.

# 3. Cybersecurity of Smart Local Energy Systems

The added ICT dimension to the classical power grid, introduced new security issues and challenges that have not been well addressed on the classical power grid. These security issues and challenges could hinder the rapid deployment and adoption by end-users of the IoT-based smart grid and future SLES.

## 3.1. Cybersecurity Objectives and Requirements

According to National Institute of Standards and Technologies (NIST), there are three cybersecurity objectives to protect information being stolen, compromised or attacked. The objectives include confidentiality, integrity and availability, also known as the CIA triads [27]. In most IT systems, confidentiality has been considered being of the greatest importance.

However, in SLES, the priority is to ensure the availability of system and secure energy supply and integrity is the next important security objective followed by confidentiality.

***Availability*** is to ensure the information is available when authorised users need to access it. In traditional power gird, utilities use limited information to estimate meter readings and hence the data availability is unlikely to cause serious impacts on the grid. However, in the SLES, the safe operation of systems is heavily reliant upon the real-time and near real-time data from the sensors across the SLES, AMI and control signals exchanged between multiple entities. As mentioned in previous section, the application of AMI not only provides consumer's data with higher resolution, but also transmits outage alarms and manage critical functions, such as distribution automation. Availability is therefore the primary security objective in the SLES to ensure the timely transmission of data, even when the network is under attack and flooded traffic [28]. The availability of data in the SLES needs a secure collection of network layers, including application layer, transport layer, network layer and physical layer [29]. Any threats or attacks on single or multiple layers in the network may cause unwanted disruptions on the communication and operations. Therefore, availability is considered as the priority in the SLES.

***Integrity*** aims to protect data and keep it in a correct state from any accidental or malicious modification of data [27]. The data must not be changed in an unauthorised or undetectable manner. It involves maintaining the consistency, accuracy and trustworthiness of data during storage, transmission, and usage. In the context of SLES, the data integrity can be targeted by attackers who attempt to alternate critical data such as metre reading, billing information and control demand. Therefore, authentication, certification and attestation are commonly adopted as protection measures [30]. The components in the SLES needs to authenticate each other so that impersonation can be detected and avoided. Then the data certificate keeps the message exchanges from any alternation and changes during the data transmission. Substations use attestation to confirm that the memory contents on a smart device have not been changed. In general, public-key cryptography are used to ensure the integrity and the services are usually provided by a trusted third party hosting a key management service.

***Confidentiality*** refers to protecting personal privacy and proprietary information from unauthorised access [27]. It emphasises the need for information protection, requiring relevant measures to ensure only authorised people being allowed to obtain the information. Attacks targeting confidentiality do not necessarily cause substantial impacts on the system operation but can be a preparatory step to more damaging attacks. With the rollout of smart meters, some concerns regarding the consumers' privacy have also been addressed in recent years [31–33] . Customers fear that the data leakage may potentially be used by unauthorised parties or marketing firms. In the future, a SLES will involve with frequent interaction between consumers, network operators and local authorities comparing to the traditional energy system, which will make confidentiality one of the most prominent issues.

### 3.2. The Development of Cybersecurity Standards

Although there are not sufficient studies and guidelines for SLES, similarities to the smart grids are helpful to understand the potential cybersecurity requirements that may be needed in a SLES. Therefore, this section aims to review the existing security standards and protocols related and provide an insight on how they can be applied in the context of SLES. There are around 100 existing standards addressing the cyber security issues, but this section we only review some of them that reflecting the main trend of the development in the chronological order. The full list of standards reviewed in the research can be found in the appendix.

The development of cybersecurity standards mirrors the trend of technological advancement. Initially, there were not any intelligent or smart control systems or devices in the system. Therefore, in the 1990s the cybersecurity protection was mainly addressed by enhancing the security of the physical assets. For instance, *IEEE 1264 Guide for Animal Deterrents for Electric Power Supply Substations* was proposed [34]. It defines types of intrusions and identified subsequent problems and impacts, evaluated by several parameters, such as intrusion location and seriousness of impacts. Correspondingly, relevant precaution and prevention measures are provided, such as physical obstacles and enclosure, security patrol and video surveillance. Later in 2000, *IEEE 1402 Guide for Electric Power Substation Physical and Electronic Security* [35] was introduced and complemented the security protection from human intrusion. Besides existing measures against animal intrusion, IEEE 1402 also includes protection measures for electronics, such as virus scans, encrypting and dial-back verification. However, these protective measures are only briefly mentioned without details in depth.

With the development of cybersecurity protection, most standards focused on the design and management at a macro systematic level or a domain-specific level. *DHS Cyber Security Procurement Language for Control System* [36] is an important documentation that combines many requirements into 11 high-level topics, such as system design, account and access control, end device management, physical and cyber threat and vulnerability detection. Each topic addresses a specific issue or concern in a control system, and also describes a rationale, from specification language to factory and site acceptance test measures. In addition, *NISTIR 7268 Guideline or Smart Grid Cyber Security* [37] specifies a comprehensive framework for smart grid. The guideline includes five steps: use case selection, risk assessment, boundaries setting, proposing security requirements, and testing and certification. The NISTIR 7268 emphasises four high-priority challenges: more cost-effective and secure devices, more advanced cryptography and key management, more robust system, and easier networking. North American Electric Reliability Corporation Critical Infrastructure Protection (NERC CIP) [38] standards are compulsory for the whole electric system. The NERC CIP standard series highlight the importance of cybersecurity of the assets that should undertake regular security risk and vulnerability assessment and equip with mandatory minimum security management controls and recovery plans. It is also worth mentioning that NERC CIP emphasises the significance of training authorised personnel with a sufficient cybersecurity awareness. In the UK, The British Standards Institution (BSI) issued *PAS 555 Cyber security risk - Governance and management – Specification* in 2013 [39]. It provides several guidelines on cybersecurity and describes the outcomes as the result of good cybersecurity practice. PAS 555 applies to all types of organisations and businesses, including the energy industry. It considers not only the technical aspects of cyber security, but also the physical, cultural and behavioural aspects, alongside effective leadership and governance.

The protection of communication network as part of industrial control and automation system was also addressed due to the proliferation of the Internet. In 2014, *IEEE C37.240*

*Standard Cybersecurity Requirements for Substation Automation, Protection, and Control Systems* [40] was developed and published to present a set of baseline cybersecurity requirements dedicated to the communication system. It aims to protect the security of interface between control systems and standardise the foundational requirements for communication components, such as access and use control, data integrity and confidentiality, network resource availability and timely response to events. Most importantly, it reemphasises the importance of monitoring and auditing security events and policies, and also conducting periodic security tests. Another instance is *IEC 62443 Industrial communication networks - IT security for networks and systems*, launched in 2010 [41]. It provides a detailed description regarding the elements and the development process of a cybersecurity management system for a control system and automaton technology. It also lists seven requirements, overlapped with the foundational requirements in IEEE C37.240, to achieve higher security levels. The security levels are considered as the functional requirements for the system, protecting from accidental information disclosure and unauthorised access and different level varies with the attacking method and activeness used by the attackers.

The increasing complex electronic devices implemented then shifted the focus towards the specific component or technology in the system. *IEEE 1686 Standard for Intelligent Electronic Devices Cyber Security Capabilities* was published in 2013 [42]. The standard series describes various compulsory requirements for the electronic devices, such as an interface to change user accounts, keeping full sequential audit history, and monitoring security-related events. The standard also requires that all electric devices should have certain level of cryptographic features to ensure the device functionality and secure communication. Another example is *Advanced Metering Infrastructure System Security Requirement* [43] issued in 2008 in the US. It aims to provide a set of security requirements to ensure the high level of information assurance, availability and security necessary to maintain a reliable system and consumer confidence. The requirements in the document can be generalised into three categories: **a) primary security services** (aims to protect confidentiality and privacy, integrity, availability, identification, authentication and authorization); **b) supporting security services** (such as detection, risk assessment, cryptography and certificate); and c) **assurance services** (such as accountability, and access control). Similar standards can also be found, such as *CEN/CLC/JTC 13 N 468 Protection Profile for Smart Meter* [44] in the UK and Privacy and *Security of the Advanced Metering Infrastructure* [45] in Netherlands.

The surge of electronic devices has markedly facilitated the digitalisation of energy system that needs to handle with substantial amount of information and data exchange. Therefore, joint efforts by academia and industry have been trying to propose relevant standards or protocols to ensure the data and information security. *IEEE 1363 Standard Specifications for Public-Key Cryptography* was firstly introduced in 2004 [46], aiming to produce a comprehensive document defining a range of common public-key techniques, including key agreement, public-key encryption and digital signatures. It describes different types of cryptographic techniques such as identity-based, password-based and lattice-based techniques and discusses relevant security and implementation considerations. *IEEE P1912 Standard for Privacy and Security Framework for Consumer Wireless Devices* [47] also focuses on data privacy and security. It provides a definition of private data, which can refer to personal identifiable information. Different types of privacy data are recommended to apply relevant necessary data settings, which is of great importance for the future applications at the end-user side. Meanwhile, more risk management guidelines are introduced to enhance the information security. For example, *ISO/IEC TR 27019 Information technology — Security techniques — Information security management guidelines based on ISO/IEC 27002 for process control systems specific to the energy utility industry* [48]. It suggests that security requirements analysis and a complementary individual risk analysis should be undertaken before the use of control devices or software. IEEE 2144.1 *Standard for Cryptographic Protection of Data on Blockchain-Oriented Storage Devices* [49] also presents a trusted IoT data management framework integrate with application, function and trusted carrier layers. The framework is applicable for data management in blockchain and IoT technologies and to business scenarios that employing internal data collection, change and sharing with external parties. *IEEE Std 11073-40101 Cybersecurity—Processes for vulnerability assessment* [50] proposing an auditable approach to identification and assessment of cybersecurity vulnerabilities and estimation of risks, which is an useful tool and can be used as reference method for future smart devices development.

# 4. Discussion and Suggestion

## 4.1. Findings in the Review of Existing Standards

The previous section has reviewed the development of cybersecurity standards and protocols that were defined and specified by industry and standard bodies. Many of them were developed to address security and privacy concerns and requirements in either control and wireless system and devices, or management strategy of cybersecurity issues. The requirements included in the standards differ from each other, in terms of technical details, the scope and the thematic coverage. It is also worth noting that some standards extend or partially repeat requirements from other standards, and some are only supplementary documents to others.

Our findings suggest that there is a considerable number of standards or protocols pre-existing that would apply to the application and infrastructure of SLES, such as industry automated and control system [51], electric vehicles [52] , and intelligent electronic devices [53]. Many standard bodies from different countries have contributed to the knowledge, such as BSI from the UK, IEEE and ANSI from the US. These standards are applicable to some specific component or industry of the infrastructure. For example, the majority of the standards proposed by IEEE defines and specifies very detailed technical security elements for the IoT. The works are applicable to many aspects and components in the SLES, including IoT architectural frameworks, physical and medium access control and wireless devices with end-to-end security.

Another finding is that the standards are not comprehensive and some only address cybersecurity to some certain extent. The existing standards are highly fragmented that are specific to certain industry, such as NERC CIP for electric utility and IEC 62443 for general industry automation and control systems, while some security frameworks providing general guidelines applicable to any industry or organisation without technical details, such as PAS

555. Majority of the standards only focus on securing one or a few components or security features in the system by the design or for the operation. Researchers have already addressed the ambiguous definition still hinders the application of the smart city [54]. Therefore, the lack of comprehensive understanding of SLES makes it even harder to provide a set of comprehensive guidelines on cybersecurity for SLES, because of the differences between SLES and smart cities, such as more localised governance bodies and more active prosumers' participation in energy supply. SLES will be heavily reliant upon substantial integration of IoT technologies and automated control and communication networks, which will make cybersecurity one of the primary goals. However, it would depend on clear definitions of SLES and explicit guidelines on its operation and governance. Therefore, relevant necessary and essential measures can be adopted to construct a robust and sophisticated smart local energy network in a more systematic fashion.

In addition, our findings also suggest that the information security is becoming increasingly important due to the growing penetration of IoT and digitalisation of the energy industry. Different standards were put in place to standardise the data encryption, transmission, storage and format to enhance the interoperability between different system components. Additionally, relevant standards were also designed to protect personal data and privacy via both algorithm and edge device design. The complex requirements are introduced to ensure certain levels of security measures embedded in the electronic devices by the manufacturers to protect the cybersecurity of both the system and users. However, HP conducted an assessment on 10 common IoT devices and they found each device had 25 vulnerabilities on average [55]. For this reason, we would recommend an adherence to certain standards must become as the norm of smart device development and sufficient and comprehensive standards are needed to be considered as the baseline standards that provides principles to ensure the scalability and flexible interpretability, which is also in line with the suggestions provided in [26].

For this reason, the findings from the review can help us understand how the existing standards can integrate with the SLES. For example, the substantial amount of data will be collected and exchanged during the daily operation, which makes the data safety and privacy protection extremely important. The potential peer-to-peer energy trading within the SLES can make the best use of the blockchain technologies and existing relevant standards such as *IEEE 2144.1* that ensure the data security in the devices and entities. Furthermore, standards similar to *PAS 555* can help us comprehend the critical role of cybersecurity and build an effective framework to assess and manage potential cyber threat, vulnerability and attacks. As the result, a more active cyber threat prevention and detection mechanism will be added to the existing protection measure, further enhancing the cybersecurity of the SLES.

## 4.2. Cybersecurity Suggestions for SLES Planning and Deployment

In the light of the emerging SLES, energy utilities have been exposed to ever substantial challenges, especially cyber challenges, which may cause catastrophic impacts on the whole value chain of the energy network. The legacy generation systems and clean-energy infrastructure without sufficient security design will likely suffer serious disruption of service and ransomware attacks against generation assets. The physical security weaknesses may possibly lead to unauthorised access to the grid control system and hence result in large-scale disruption of power to customers through remotely disconnecting services. At distribution level, limited security measures built into SCADA systems can cause disruption of reginal loss and disruption of service to customers. At network and end-user side, large attack surface of IoT devices such as smart meters, electric vehicles and other smart appliances, will also possibly lead to theft of customer information, fraud, and service disruption.

The SLES aims to achieve an automated and local energy supply with high participation of prosumers with the help of highly penetrated IoT technologies. The difference in the operation and management of SLES and traditional power grid will contribute to the merging of information and operation technologies. For this reason, the challenges are brand-new and unprecedent, and can hardly be solved by using traditional cyber threat management strategy. Therefore, the researchers from academia and industry need to work on several things and we have made following recommendations.

Due to interdependency between the physical and cyber infrastructures, the cybersecurity of SLES should focus on protecting measures on both physical and cyber aspects. It is hard to detail every cybersecurity requirement here, but a good comprehensive cybersecurity guideline should include following 15 aspects: access control, audit and accountability, configuration management, identification and authentication, incident response, media protection, planning, personnel security, information system and service acquisition and integrity, awareness and training, security assessment and authorisation, information and document management, physical and environmental security, risk assessment and management, and communication system protection.

In addition, for the industry, more efforts are needed to provide more affordable and cyber-secured devices and services and more innovative technologies should also be encouraged to apply in real applications. Technologies, such as blockchain and OpenFMB, can not only facilitate scaling up the SLES applications with secured assurance, but also can improve the integration with legacy infrastructures with enhanced data interoperability. A good balance between the affordability and quality of cybersecurity should be achieved so that IoT products can be more easily accessed by consumers. In comparison, the government and regulator should introduce clear guidelines or regulatory documents on baselines cybersecurity requirements and other relevant issues on data management. Therefore, clear definitions and guidelines on SLES should be considered as priority. A tailored cybersecurity management strategy needs to be designed based upon good comprehension of a system setup, operation, and governance. A few baseline standards are needed to address the system's baseline security requirements, so that relevant components or technologies can then be adopted to meet the minimum function and security requirements.

At last, compliance testing and certifications should also play an important role in SLES. Although there may be technical standards already in place to ensure the security at the design or development stage, the consistency should come across the whole SLES ecosystem. For this reason, it is necessary to conduct testing and certification by an independent party, which can assure the regulators that a satisfied security level is provided in key SLES ecosystem actors by design.

## 5. Conclusion

In this report, a review on existing technical standards addressing cybersecurity issues is carried out. Our findings suggest that a considerable number of standards or protocols pre-existing that would meet the requirements of the application and infrastructure of SLES. However, the standards are not comprehensive and some only address cybersecurity to some certain extent. The existing standards are highly fragmented that are specific to certain industry, while some security frameworks providing general guidelines applicable to any industry or organisation without technical details. Majority of the standards only focus on securing one or a few components or security features in the system by the design or for the operation. Additionally, we also find that the information security is becoming increasingly important, and many standards are introduced to protect information security and personal privacy.

However, the successful development of SLES still needs more effort from multiple sides. A detailed cybersecurity guideline should include 15 main topics described in previous section. More efforts are needed from the industry to provide more affordable and cyber-secured devices and services to apply in real applications. The governments and regulators should demonstrate a few baseline standards to address the system's baseline security requirements, so that relevant components or technologies can therefore be adopted to meet the minimum function and security requirements. Additionally, compliance testing and certifications should also be in place and carried out by an independent third party to ensure the components of SLES ecosystem with a satisfied security level by design.

Based on our findings and suggestions produced from this research, it is important to extend the research and further investigate how they can contribute to the design and operation of SLES. The future work will focus on proposing a framework for the detection and treatment to ensure the cybersecurity of SLES. The first part of the framework results in an assessment tool that aims to identify the potential cybersecurity vulnerabilities, threats, and attacks. The evaluation on the threats will be determined via various metrics, such as the type of threats, location in the cyber and physical architecture, and responsible stakeholders. The assessment results will then result in a strategic response. The threat metrics are used to prioritise the tasks based on the urgency and seriousness level of the threat. Relevant solutions and techniques are then provided to respond and prevent the system from potential threats throughout the cyber, physical devices or utility layers to ensure the cyber-secured operation.

## Acknowledgement

This work was partially supported by the EPSRC EnergyREV project (EP/S031863/1). We would like to thank Elena Gaura, Nandor Verba, David Flynn and Rameez Asif for helpful insights and expertise that greatly assisted the research.

## Appendix: Cybersecurity Standards Reviewed in the Study

| Standard | Year | Description | Coverage | Purpose | Highlights |
|---|---|---|---|---|---|
| IEEE Std 1264 [34] | 1993 | IEEE Guide for Animal Deterrents for Electric Power Supply Substations | Substation, Nonhuman intrusion | Counteract animals' intrusions | 1) Define type of animal intrusions and possible problems; 2) Precaution and prevention measures, such as barriers and enclosures |
| IEEE Std 1402 [35] | 2000 | IEEE Guide for Electric Power Substation Physical and Electronic Security | Substation, human intrusion | Counteract human intrusions | 1) Define type of intrusions, including pedestrian, vehicular, projectile, electronic; 2) Define parameters and events influencing intrusions and subsequent problems; 3) Security criteria and management are proposed. |
| IEEE 802.1AE [56] | 2006 | Standard for Local and Metropolitan Area Networks: Media Access Control (MAC) Security | Metropolitan Area Networks | Specifying provision of connectionless user data confidentiality, frame data integrity, and data origin authenticity | Specifies 1) requirements that devices need to comply with; 2) requirements for and definitions of MAC security; 3) management strategy of MAC security. |
| ANSI C12.18 [57] | 2006 | Protocol Specification for ANSI Type 2 Optical Port | Communication between end-device and clients via optical port | To detail the criteria for the communication and details for implementing OSI 7-layer model. | Specify the use of ANSI type 2 optical port for meter communications |
| ANSI C12.21 [58] | 2006 | American National Standard for Protocol Specification for Telephone Modem Communication | Utility communication over telephone modem | To specify requirements for communication amongst users and devices over switched telephone modem | Specify the use of telephone modem for meter communications |
| IEEE std 1619 [59] | 2008 | Standard for Cryptographic Protection of Data on Block-Oriented Storage Devices | Data encryption in sector-based devices | Providing a cryptographic protection for data stored in constant length blocks. | 1) Defines specific of an architecture for cryptographically protecting data stored in constant length blocks; 2) provides an additional and improved tool for implementation of secure and interoperable protection of data residing in storage. |
| IEEE std 2600 [60] | 2008 | Standard for Information Technology: Hardcopy Device and System Security | Security requirements for hard-copy devices | Providing a guidance on security requirements for hard copy devices during manufacture, installation, and applications. | 1) To provide guidance in the secure architecture, design, and out-of-box configuration of HCDs for manufacturers; 2) To provide guidance in the secure installation, configuration, and use of HCDs for end users and their supporting organizations. |
| ANSI C12.19 [61] | 2008 | American National Standard for Utility Industry End Device Data Tables | Utility data table structure | Defining structures for transporting data to and from end devices | Defines a table structure for utility application data to be passed between an end device and a computer. Does not define device design criteria nor specify the language or protocol used to transport that data. |
| ANSI C12.22 [62] | 2012 | American National Standard for Protocol Specification for Interfacing to Data Communication Networks | Utility data interoperability and security | Specify the transportation of data over various networks to advance interoperability and security amongst communication modules and meters. | 1) Describes the process of transporting C12.19 table data over a variety of networks; 2) Uses AES encryption to enable strong, secure Smart Grid communications, including confidentiality and data integrity, and is also fully extensible to support additional security mechanisms the industry may require in the future. |
| IEEE 802.21a [56] | 2012 | Standard for Local and Metropolitan Area Networks: Media Independent Handover Services - Amendment for Security Extensions to Media Independent Handover Services and Protocol | Metropolitan Area Networks | Protecting media independent handover services and mechanisms; Assisting proactive authentication to reduce the latency due to media access authentication and key establishment with the target network. | 1) to protect MIH messages, (D)TLS based protection when a PKI involved and EAP based authentication are introduced. 2) to reduce the latency, three key distribution mechanisms are introduced. |
| IEEE Std 1703 [63] | 2012 | IEEE Standard for Local Area Network/Wide Area Network (LAN/WAN) Node Communication Protocol to Complement the Utility Industry End Device Data Tables | Interoperability among communications modules and meters | Accommodating the network messaging requirements of an advanced metering infrastructure | 1) it uses advanced encryption standard to enable strong secure communications to protect confidentiality and data integrity. 2) the security model is extensible to support new security mechanism, but the cipher model cannot secret non-standard short messages. |
| IEEE 1686 [53] | 2013 | IEEE Standard for Intelligent Electronic Devices Cyber Security Capabilities | IED | Addresses security regarding the access, operation, configuration, firmware revision and data retrieval from an IED, and communication encryption. | 1) IEDs should have open and documented interface to change user accounts, passwords, and roles. 2) IEDs should keep full sequential record of audit history (at least 2048 events). 3) IEDs should monitor security-related activities and inform SCADA through a real-time protocol. 4) IEDs should have certain cryptographic features to ensure the communication and functionality with the help of various techniques. 5) Firmware quality assurance shall follow IEEE std C37.231 |
| IEEE Std 1363.3 [64] | 2013 | Standard for Identity-Based Cryptographic Techniques using Pairings | Identity-based cryptographic schemes based on the bilinear mappings over elliptic curve | Specify identity-based cryptographic techniques based on pairings. | Techniques for identity-based encryption, signatures, inscription, key agreement, and proxy re-encryption, all based on bilinear pairings. |
| PAS 555 [39] | 2013 | Cyber security risk. Governance and management. Specification | A business-led, holistic approach to cyber security | Define the overall outcomes of effective cyber security | PAS 555 enables any organization to choose how it achieves the specified outcomes, whether through its own defined processes or the adoption of other standards and management systems. PAS 555 enables organizations to 1) Focus investment in the most appropriate way; 2) Minimize potential loss; 3) Improve operational effectiveness and efficiency; 4) Develop organizational resilience; 5) Improve loss prevention and incident management; 6) Identify and mitigate cyber security risk throughout the organization |
| TIA TSB-4940 [65] | 2013 | Smart Device Communications; Security Aspects | ICT operation security | Addressing only the management of cyber security related risk derived from or associated with the operation and use of information technology | Provide a framework that can protect communication security, including assessing external threat, vulnerability assets, and relevant approach to protect vulnerable assets. |
| ISO/IEC 27001 -27005 [48] | 2013 | Information technology — Security techniques — Information security risk management | Information security risk management | Assisting the satisfactory implementation of information security in all types of organisations based on a risk management approach. | It specifies how to 1) identify and assess the risks; 2) deal with the risks; 3) monitor the risks and risk treatments; 4) Keep stakeholders informed throughout the process. |
| IEEE C37.240 [40] | 2014 | IEEE Standard Cybersecurity Requirements for Substation Automation, Protection, and Control Systems | Substation, interfaces | Minimum requirements for a substation program. Any utility's CS programme must balance technical, economic, and operational feasibility. | 1) Foundational requirements: access control, use control, data integrity, data confidentiality, restrict data flow, timely response to event, network work recourse availability. 2) Physical Security needs to ensure the secure access to the cyber assets. 3) Protection of data at the rest including file-type of IEDs (configuration files, data files, etc) and hard copy information (IED instruction manuals, station drawings) |
| IEEE 2410 [66] | 2015 | Standard for Biometric Open Protocol | Biometric Open Protocol Standard | The Biometric Open Protocol Standard (BOPS) provides identity assertion, role gathering, multilevel access control, assurance, and auditing. | 1) BOPS introduces the security considerations, including identity assertion, role gathering, access control, audit and assurance; 2) how BOPS is realised in applications and relevant requirements. |
| IEC 27019 [48] | 2017 | Information technology — Security techniques — Information security controls for the energy utility industry | Energy management cybersecurity | Guidance on process control systems for utility industry for controlling and monitoring the production and generation, T&D | Specifies the cybersecurity requirements for several components, including central and distributed process control, monitoring, communication technology, advanced metering infrastructure, energy management system, software, firmware and remote maintenance system |
| IEEE P2413 [67] | 2019 | Standard for an architectural framework for the Internet of Things | IoT architectural framework | Promoting cross-domain interaction and system interoperability and functional compatibility and accelerating the growth of IoT market | 1) recognizes the evolving transformational integration and convergence across technology and application domains; 2) to provide an extensible integrated architectural framework that will continue to evolve and unify the standards creation effort; 3) also provide enough flexibility for different industries to adapt the acritude based on different needs. |
| IEEE Std 11073-40101 [50] | 2020 | Cybersecurity—Processes for vulnerability assessment | Personal Health Devices and Point-of-Care Devices | Proposing an auditable approach to identification and assessment of CS vulnerabilities and estimation of risks | 1) emphasise the role of device manufacturers that should provide a device with sufficient security protection measures and without any hidden and undocumented functions; 2) Provide two good threat modelling approaches data flow diagram and STRIDE classification scheme (***S**poofing, **T**ampering, **R**epudiation, **I**nformation **D**isclosure, **D**enial of Service and **E**levation of Privilege*) 3) A scoring system is proposed to quantify vulnerabilities, which provides a rank and priority to each vulnerability. |
| IEEE 2144.1 [49] | 2020 | Standard for Cryptographic Protection of Data on Blockchain-Oriented Storage Devices | IoT, Blockchain, data | Proposing a framework of blockchain-based IoT data management | 1) define the roles of stakeholders in a IoT system, including data owner, data consumer, service provider, regulator/policy maker, etc. 2) specify blockchain-based data management lifecycle; 3) proposing trusted IoT data management framework integrate with 3 layers: application, function and trusted carrier layers, which all communicated and controlled by management and control panel. |
| IEC 61969 [68] | 2011-2020 | Mechanical structures for electrical and electronic equipment - Outdoor enclosures | Design standard for outdoor enclosure | Establish basic environmental performance criteria for outdoor enclosure compliance | Defining design guidelines for outdoor enclosures and is applicable over mechanical, electromechanical and electronic equipment and its installation |
| IEC 61970 [69] | 2005-2020 | Energy management system application program interface | Common Information Model for transmission network domain | Standardising the data format and enhance interoperability at transmission network level | 301 contains a standard data model defining the semantics of the information exchanged in a broad range of energy management system applications and later version 302 intended to ensure the data interoperability among transient stability software products. Later versions such as 45x, 452 and 453 tend to standardise the data profiles for state estimation and diagram layout profiles for data exchange. |
| ISO/SAE 21434 [52] | 2020 | Road vehicles - Cybersecurity engineering | Road Vehicle | Proposing a technical standard for automotive development that can demonstrate compliance with regulations in EU. | The standard provides guidelines on how to manage cybersecurity strategies in different stages, including: overall management, during concept phase, during product development, and during production, operation and maintenance. |
| IEEE P1912 [47] | 2021 | Standard for Privacy and Security Framework for Consumer Wireless Devices | Privacy and security, consumer wireless device | Data privacy and security at end-user side | 1) Define a privacy scale applied to data collected, processed or shared amongst services or network. 2) the input of privacy data contributes to assessment tools to apply relevant necessary setting to the data. |
| DHS cybersecurity [70] | 2021 | Cybersecurity Requirements for Critical Pipeline Owners and Operators | Pipeline owners and operators | Details basic requirements for pipeline owners and operators | 1) Owners and operators must report confirmed and potential cybersecurity incidents. 2) Owners and operators must designate a Cybersecurity Coordinator to be available 24 hours a day, seven days a week.; 3) Owners and operators must review current practices to identify and remediate gaps related to cyber risk and report all findings to both TSA and CISA within 30 days. |
| IEEE Std 1363 series [46,64,71] | 2004-2008 | IEEE Standard Specifications for Public-Key Cryptography | public-key cryptography techniques | Electronic privacy and authenticity | Introducing types of cryptographic techniques and speciation's of key agreement schemes, signature schemes, encryption schemes. |
| ISO 16484 [72] | 2004-2020 | Building Automation Controls network | Communication across devices | Enhancing interoperability amongst different vendors and equipment | 1) Specify guiding principles for project design and implementation and integration of other systems into automation and control system; 2) specifies the requirements for the hardware to perform the tasks; 3) specifies the reequipments for overall functionality and engineering service; 4) defines data communication services and protocols for computer equipment used for building systems. |
| IEC 62443 [41] | 2009-2020 | Industrial communication networks - IT security for networks and systems | Industrial and automation systems cybersecurity | Addressing and mitigating current and future security vulnerabilities in industrial automation and control systems | 1) Specifies process requirements for the secure development of products used in an IACS and defines a secure development life cycle for developing and maintaining secure products; 2) Details the cybersecurity technical requirements for components making up an industrial automation and control system, including embedded devices, network, host and software applications; 3) specifies security capabilities enabling a component to mitigate threats for a given security level without compensating countermeasures; |
| NIST SP 800-82 [51] | 2011-2021 | Guide to Industrial Control Systems (ICS) Security | Industrial control system security | Securing industrial control system, while addressing their unique performance, reliability, and safety requirements | The document provides an overview of ICS and typical system topologies, identifies typical threats and vulnerabilities to these systems, and provides recommended security countermeasures to mitigate the associated risks |
| IEC 61850 [73] | 2009-2021 | Communication networks and systems for power utility automation | Communication network, Substation | Specifying communication network at substation | 1) Using a data and object model to specify data format in system and devices; 2) establishing multi-vendor interoperability |